\documentclass{iau-FM}

\usepackage{mathrsfs}
\usepackage{amsfonts}
\usepackage{amstext}
\usepackage{amsmath}
\usepackage{graphics,graphicx}
\usepackage{multirow}
\usepackage{natbib}
\usepackage{enumerate}

\title[The empirical laws of galaxy dynamics]{The empirical laws of galaxy dynamics: \\ from gas kinematics to weak lensing}

\author[F. Lelli et al.]{Federico Lelli$^1$, Tobias Mistele$^2$, Stacy S. McGaugh$^2$, James M. Schombert$^{3}$, and Pengfei Li$^{4}$}
\affiliation{$^1$Arcetri Astrophysical Observatory, INAF, Largo Enrico Fermi 5, 50125, Florence, Italy \\
$^{2}$Case Western Reserve University, 10900 Euclid Avenue, Cleveland, Ohio 44106, USA \\
$^{3}$Department of Physics, University of Oregon, 1371 E 13th Ave, Eugene, Oregon 97403, USA \\
$^{4}$School of Astronomy and Space Science, Nanjing University, Nanjing, Jiangsu 210023, China}

\pubyear{2024}
\jname{Astronomy in Focus, Focus Meeting 9} 
\editors{Diana M.~Worrall, ed.}

\begin{document}

\maketitle

\begin{abstract}
Galaxies obey a set of strict dynamical laws, which imply a close coupling between the visible matter (stars and gas) and the observed dynamics (set by dark matter in the standard cosmological context). Here we review recent results from weak gravitational lensing, which allow studying the empirical laws of galaxy dynamics out to exceedingly large radii in both late-type galaxies (LTGs) and early-type galaxies (ETGs). We focus on three laws: (1) the circular velocity curves of both LTGs and ETGs remain indefinitely flat out to several hundreds of kpc; (2) the same baryonic Tully-Fisher relation is followed by LTGs and ETGs; (3) the same radial acceleration relation (RAR) is followed by LTGs and ETGs. Combining galaxy data with Solar System data, the RAR covers about 16 orders of magnitude in the Newtonian baryonic acceleration. Remarkably, these empirical facts were predicted a priori by MOND.
\end{abstract}

%\begin{keywords}
%Dark matter --- galaxies: kinematics and dynamics --- galaxies: elliptical and lenticular, cD --- galaxies: spiral --- gravitation --- gravitational lensing
%\end{keywords}

\vspace{-0.5cm}
\section{Introduction}

Galaxies display remarkable regularities in their dynamical properties, which can be summarized by a basic set of empirical laws:
\begin{enumerate}[1.\,]
  \item \textbf{Flat Rotation Curves.} The circular velocity of a galaxy reaches an approximately constant value that persists indefinitely at large radii \citep{Rubin1978, Bosma1978}.
 \item \textbf{Renzo's Rule.} For any feature in the circular-velocity curve of a galaxy, there is a corresponding feature in the baryonic density profile, and vice versa \citep{Sancisi2004}.
  \item \textbf{Baryonic Tully-Fisher Relation (BTFR).} At large radii, the ``flat'' circular speed correlates with the total baryonic mass \citep{McGaugh2000, Lelli2019}.
  \item \textbf{Central Density Relation (CDR).} At small radii, the central dynamical surface density correlates with the central baryonic surface density \citep{Lelli2013, Lelli2016c}.
  \item \textbf{Radial Acceleration Relation (RAR).} At each radius, the observed acceleration correlates with the baryonic acceleration \citep{McGaugh2016, Lelli2017}.
\end{enumerate}
The term ``circular velocity'' ($V_{\rm c}$) refers to the rotation speed of a test particle on a circular orbit under the gravitational potential ($\Phi$), so that $V_{\rm c}^2 = - R\cdot\partial_{\rm R} \Phi$ where $R$ is a cylindrical radius. In \citet{McGaugh2020a} and \citet{Lelli2022}, we reviewed these laws focusing on kinematic data. In this review, after recalling some basic facts, we focus on weak gravitational lensing data.

In the particle dark matter (DM) context, each one of the dynamical laws implies a different type of baryon-DM coupling and/or fine-tuning problem in galaxy formation:
\begin{enumerate}[1.\,]
 \item The asymptotic flatness of rotation curves implies that the radially-declining baryonic contribution ($V_{\rm bar}$) and the radially-rising DM contribution ($V_{\rm DM}$) must be fine-tuned in order to have $V_{\rm bar}(R)^2 + V_{\rm DM}(R)^2 = const$ at each $R$. This is historically referred to as ``disk-halo conspiracy'' \citep{vanAlbada1986} and remains a problem for $\Lambda$CDM models of galaxy formation \citep{Desmond2017b, Desmond2019}.
 \item Renzo's rule implies that local variations in the baryonic mass distribution correspond to local variations in the total gravitational potential, even when DM supposedly dominates, so that baryons and DM must be locally coupled \citep{Sancisi2004}.
 \item The BTFR implies that the \emph{global} baryonic-to-DM mass ratio ($f_{\rm bar} = M_{\rm bar}/M_{\rm DM}$) must systematically vary across galaxies with virtually no intrinsic scatter at a given mass, despite the stochastic process of galaxy formation \citep[e.g.,][]{Desmond2017b}. In addition, the lack of redisidual correlations with mean baryonic surface density ($\Sigma_{\rm bar}$) requires that $\Sigma_{\rm bar}\cdot f_{\rm bar}=const$ at fixed mass, which is another fine-tuning problem. 
 \item The CDR implies that the \emph{central} baryonic-to-DM ratio in galaxies must systematically vary with $\Sigma_{\rm b}(0)$, with a characteristic ``break'' below which DM starts to dominate. In addition, the lack of residual correlations with total mass is puzzling because Newton's shell theorem does not hold for a disk, so the dynamical surface density at small radii should depend also on the mass distribution at large radii \citep{Lelli2016c}.
 \item The RAR implies that the \emph{local} baryonic-to-DM ratio at each $R$ systematically depends on the Newtonian baryonic acceleration ($g_{\rm bar}$), with a characteristic acceleration scale below which DM starts to dominate. In addition, the lack of residual correlations with other galaxy properties is puzzling because, in a galaxy disk, $g_{\rm bar}(R)$ is given by a complex integral of the baryonic surface density at every $R$ \citep{Casertano1983}. 
\end{enumerate}
The BTFR, CDR, and RAR imply the existence of three logically distinct acceleration scales, each one playing a different role in galaxy dynamics \citep{Lelli2022}. These acceleration scales, however, display a consistent value of $\sim$10$^{-10}$ m s$^{-2}$, suggesting a common origin.

In the context of Milgromian dynamics \citep[MOND,][]{Milgrom1983b, Milgrom1983a}, the acceleration scales are identified with a new constant of Nature, $a_0$, which sets the transition from the high-acceleration Newtonian regime to the low-acceleration Milgromian regime. Then, the dynamical laws can be derived from the basic tenets of MOND \citep{Milgrom2014a}. Actually, most of them were predicted by MOND in advance of the observations \citep{McGaugh2020b}. 

The empirical laws of galaxy dynamics have been mostly studied in late-type galaxies (LTGs; spirals and dwarf irregulars) because they usually possess a rotation-supported gas disk with negligible pressure support, so that the circular velocity is directly probed by the observed rotation speeds. Early-type galaxies (ETGs; elliptical and lenticulars) may occasionally host rotation-supported gas disks \citep{denHeijer2015}, but it is more common to measure their circular velocities by modeling the stellar kinematics \citep[considering both rotation and pressure support, e.g.,][]{Cappellari2016} or the hydrostatic equilibrium of the X-ray gas halos \citep[e.g.,][]{Buote2012}. These different methods indicate that ETGs generally follow the same dynamical laws as LTGs \citep{denHeijer2015, Lelli2017, Shelest2020}. 

In the following, we review recent results from galaxy-galaxy weak gravitational lensing, which allow us to study the dynamical laws (1), (3), and (5) in both LTGs and ETGs out to exceedingly large radii: several hundreds of kpc.

\begin{figure}[t]
\begin{center}
%  \centerline{\vbox to 6pc{\hbox to 10pc{}}}
  \includegraphics[width=\textwidth]{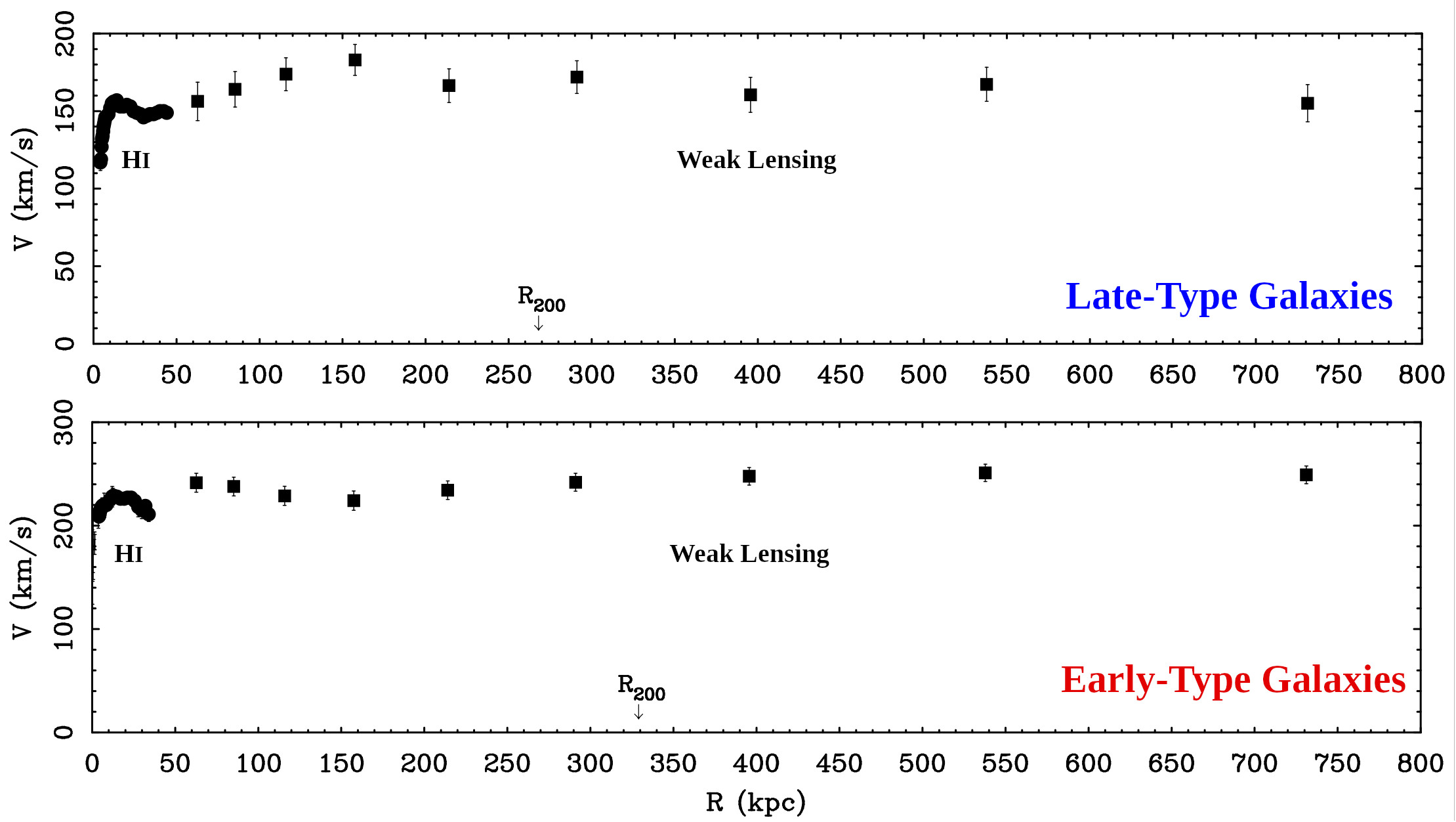}
  \caption{The circular velocities from weak lensing data (squares with errorbars) remain constant out to hundreds of kpc, well beyond the virial radius ($R_{200}$) of their expected DM halos \citep{Mistele2024b}. The same phenomenology is seen in both LTGs (top) and ETGs (bottom). For reference, the circles at $R<50$ kpc show the HI rotation curves of two individual galaxies with similar masses as the lensing ensamble: the Sc NGC\,3198 \citep[top, from][]{Begeman1989} and the S0 UGC\,6786 \citep[bottom, from][]{Noordermeer2007}.}
  \label{fig:Vc}
\end{center}
\end{figure}

\section{Galaxy dynamics from weak gravitational lensing}

\subsection{The method}

Gravitational lensing is sensitive to the total gravitational field of massive objects, so it can be used to study galaxy dynamics \citep{Brouwer2021}. Weak gravitational lensing measures little distortions induced by a ``lens'' on many background ``source'' galaxies. In practice, one measures the ellipticities of the source galaxies and infer the tangential shear $\gamma_{\rm t}$ by azimuthal averaging within a given projected radii $R_{\rm p}$ from the lens galaxy. The lensing signal induced by an individual galaxy, however, is weak, so one must stack over tens of thousands of lenses. During the stacking, the intrinsic ellipticities of the source galaxies are averaged out, so that one probes only the actual lensing distortions. Commonly, $\gamma_{\rm t}$ is used to calculate the so-called excess mass surface density ($\Delta \Sigma$).

Assuming spherical symmetry, \citet{Mistele2024a} derived a new formula to convert $\Delta \Sigma$ into the observed acceleration ($g_{\rm obs}$) produced by the total mass distribution:
\begin{equation}\label{eq:gobs}
 g_{\rm obs}(r) = \dfrac{V_{\rm c}^2(r)}{r} = 4 G_{\rm N} \int_{0}^{\pi/2} \Delta \Sigma \left( \dfrac{r}{\sin(\theta)} \right) d\theta,
\end{equation}
where $G_{\rm N}$ is Newton's constant and $r$ is the spherical radius. The average $g_{\rm obs}(r)$ can then be inferred by staking in bins of galaxy type, mass, or both \citep{Mistele2024b, Mistele2024a}.

Eq.\,\ref{eq:gobs} is expected to be valid in any metric theory where the matter fields (such as baryons and photons) are minimally coupled to the metric, so that in the quasi-static weak-field limit the metric has the same form as in General Relativity, just with a different gravitational potential than Newton's. A technical complication of Eq.\,\ref{eq:gobs} is that the integral requires measurements for $R_{\rm p}\rightarrow \infty$, so we need to extrapolate the data beyond the last available measurement of $\Delta\Sigma$. This extrapolation sets the radial range of applicability of the method. Using KiDS data \citep{Brouwer2021}, \citet{Mistele2024b, Mistele2024a} found that the extrapolation becomes significant beyond $\sim$1 Mpc, so measurements of $g_{\rm obs}$ at $r>1$ Mpc are not considered. Another technical issue is that we aim to measure the average $g_{\rm obs}(r)$ of individual galaxies, so we need to select only isolated lenses before stacking their signals. To this aim, \citet{Mistele2024b, Mistele2024a} considered only lenses that have no other galaxy with more than 10$\%$ of their stellar mass within 4 Mpc.

\subsection{Indefinitely flat rotation curves}

\begin{figure}[t]
\begin{center}
%  \centerline{\vbox to 6pc{\hbox to 10pc{}}}
  \includegraphics[width=\textwidth]{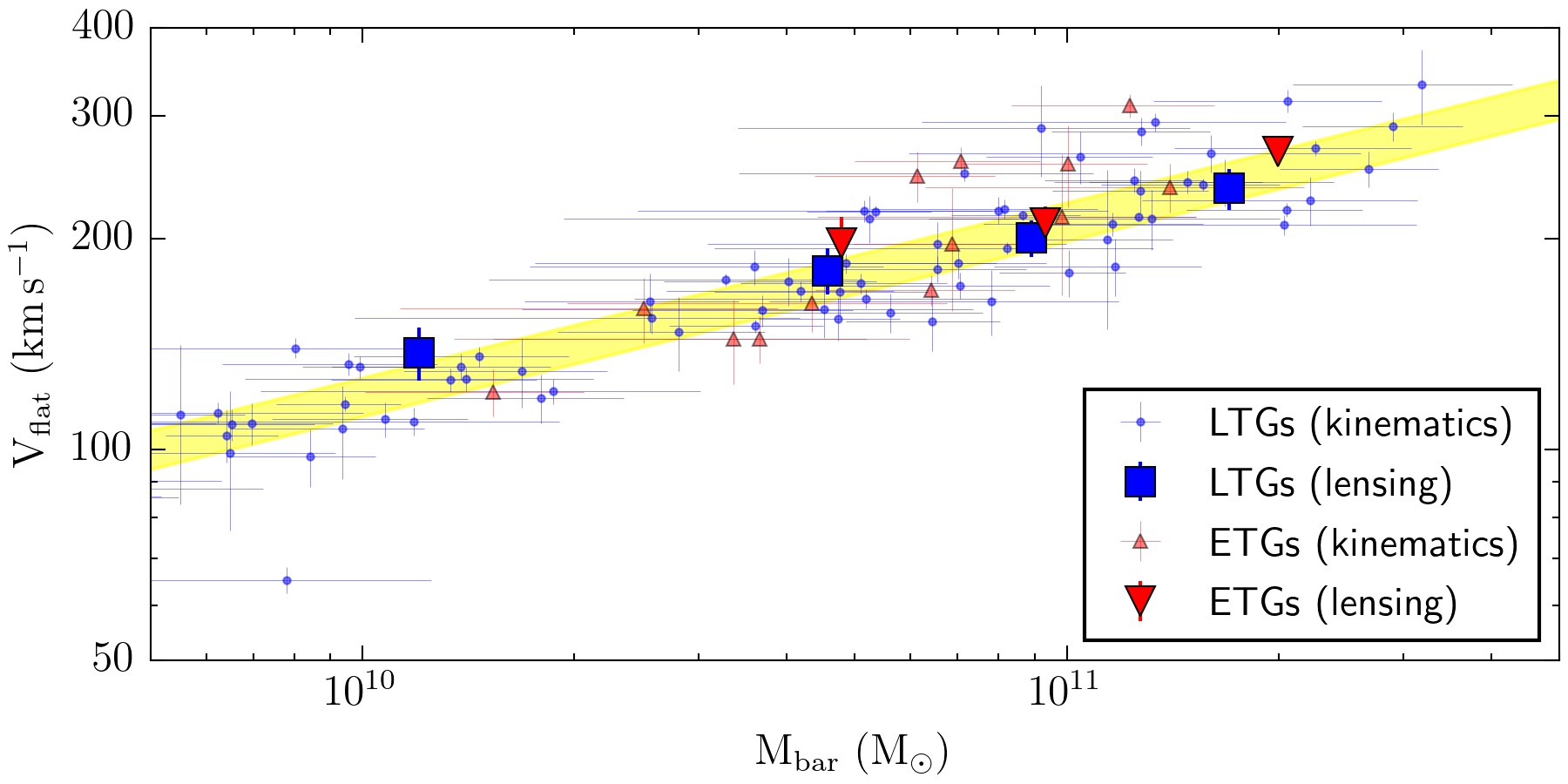}
  \caption{The BTFR focusing on the mass range $M_{\rm bar} = 5 \times 10^9 - 5 \times 10^{11}$ M$_\odot$. Large symbols show statistical weak lensing measurements for LTGs (blue squares) and ETGs (red down-triangles) from \citet{Mistele2024b}. Small symbols show individual galaxies with kinematic data; LTGs (blue circles) from \citet{Lelli2019} and ETGs (red up-triangles) from \citet{denHeijer2015}. The yellow band shows the MOND prediction considering a 25$\%$ error on $a_0$.}
  \label{fig:BTFR}
\end{center}
\end{figure}

Figure\,\ref{fig:Vc} shows the circular velocities from KiDS DR4 weak-lensing data \citep{Mistele2024b}, measured for LTGs and ETGs separately. For both galaxy types, the circular velocities remain approximately flat for several hundreds of kpc. This enormously extends the classic result from HI rotation curves \citep{Bosma1978, vanAlbada1986}.

In the $\Lambda$CDM context, the flat part of the circular velocity curves extends well beyond the expected virial radius of the DM halos, where we would expect to see a Keplerian decline \citep[see also Fig. 2 in][]{Mistele2024b}. In principle, a way to obtain such flat rotation curves in $\Lambda$CDM is to assume that the lenses are not sufficiently isolated (despite the strict isolation criterion) and that the contributions due to neighboring DM halos (the so-called "two-halo term") have the right shape and amplitude to produce a constant $V_{\rm c}(r)$. This approach raises a new ``two-halo conspiracy'' analogous to the classic ``disk-halo conspiracy'' at small radii \citep{vanAlbada1986}: the declining contribution of the one-halo term ($V_{\rm DM, 1}$) needs to conspire with the rising contribution of the two-halo term ($V_{\rm DM, 2}$) such that $V^2_{\rm DM, 1}(R) + V^2_{\rm DM, 2}(R) = const$ at large radii. In other words, the neighboring DM halos should arrange themselves around the primary halo in a fine-tuned way to make the circular velocity flat. This scenario appears quite contrived.

In the MOND context, the circular velocities of isolated systems are predicted to remain constant at large radii because the equations of motion become scale-invariant at low accelerations \citep{Milgrom2009}. For non-isolated systems, instead, MOND predicts a mild decline due to the external field effect \citep[][]{Chae2020, Chae2021}. The isolation criterion adopted by \citet{Mistele2024a, Mistele2024b} seems strict enough to avoid a significant external field effect from both nearby neighbors and the large-scale structure of the Universe.

\subsection{Baryonic Tully-Fisher relation}

Figure\,\ref{fig:BTFR} shows the BTFR considering individual galaxies with kinematic data \citep{denHeijer2015, Lelli2017, Lelli2019} together with the statistical weak lensing measurements \citep{Mistele2024b}. The two different datasets are fully consistent with each other. 

In the $\Lambda$CDM context, the BTFR must be the end result of the haphazard process of galaxy evolution. Thus, it is far from trivial to have LTGs and ETGs on the same relation because the two galaxy populations surely had different evolutionary histories (different merging histories, gas accretion histories, star-formation histories, and so on).

In the MOND context, the BTFR is a ``Natural Law'' with a fixed slope of 4. The yellow band in Fig.\,\ref{fig:BTFR} shows the MOND prediction considering current uncertainties in the value of $a_0$ \citep{Lelli2022}. The data are clearly consistent with the MOND prediction.

\begin{figure}[t]
\begin{center}
%  \centerline{\vbox to 6pc{\hbox to 10pc{}}}
  \includegraphics[width=\textwidth]{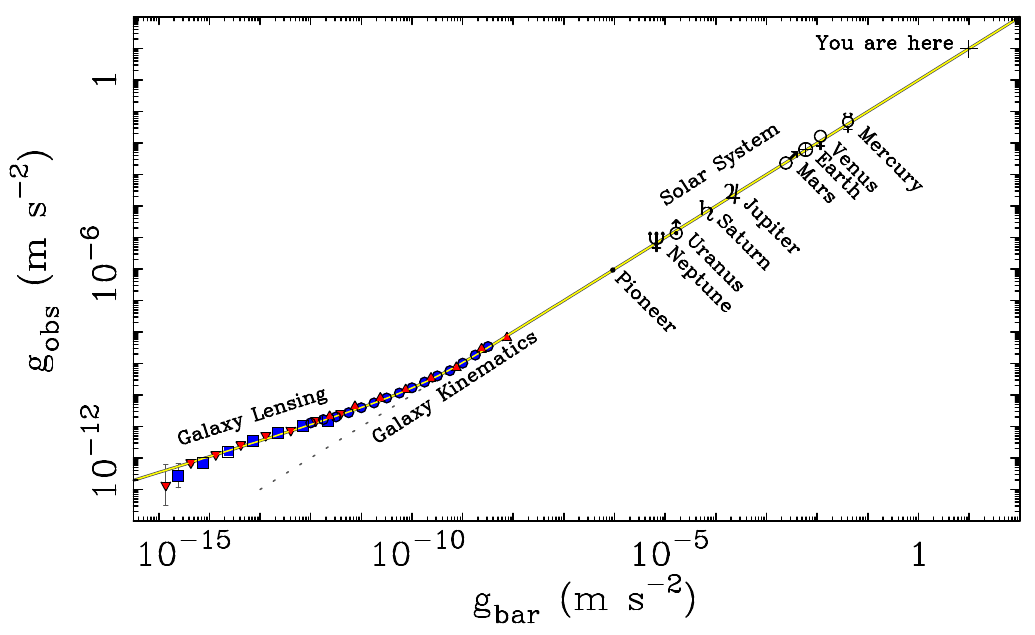}
  \caption{The RAR over the widest possible dynamic range: $\sim$16 orders of magnitude in $g_{\rm bar}$. Symbols at high accelerations (where $g_{\rm bar} = g_{\rm obs}$) show Earth's surface gravity, the mean orbital acceleration of planets in the Solar System, and the farthest measured acceleration of the Pioneer spacecraft. Symbols at low accelerations show kinematic data for LTGs (blue circles) and ETGs (red up-triangles), and lensing data for LTGs (blue squares) and ETGs (red down-triangles). At accelerations below $\sim$10$^{-10}$ m s$^{-2}$, the data display the DM effect ($g_{\rm bar} \neq g_{\rm obs}$) and follow the relation $g_{\rm obs} = \sqrt{a_0 g_{\rm bar}}$ (yellow line) predicted a-priori by MOND for isolated systems.}
  \label{fig:rar}
\end{center}
\end{figure}
\subsection{Radial Acceleration Relation}

Figure\,\ref{fig:rar} shows the RAR over the widest possible range, from the Earth's surface gravity to the outermost parts of galaxies. The kinematic data of galaxies probe the range $10^{-8}-10^{-12}$ m\,s$^{-2}$ in $g_{\rm bar}$, while weak lensing data the range $10^{-12}-10^{-15}$ m\,s$^{-2}$. Empirically, the DM effect ($g_{\rm obs} > g_{\rm bar}$) kicks in at $\sim10^{-10}$ m\,s$^{-2}$. This value is 4 orders of magnitude smaller than the lowest acceleration where Newtonian dynamics has been directly tested in non-relativistic systems, corresponding to the outermost parts of the Solar System.

The RAR has two acceleration scales: one setting the transition $g_{\rm obs} > g_{\rm bar}$ and one setting the overall normalization. Remarkably, the two acceleration scales coincide within the errors, so we have been referring to them as $g_{\dagger}$ \citep[see Sect. 3.2 in][]{Lelli2017}. The bottom portion of the RAR implies $g_{\rm obs}=\sqrt{g_\dagger g_{\rm bar}}$, which is mathematically equivalent to a BTFR with slope of 4 and no radial dependencies \citep[see Sect. 7.1 in][]{Lelli2017}.

In the $\Lambda$CDM context, the physics that may set these acceleration scales is unclear; a possibility is to assume that the star-formation efficiency of galaxies is related to the gravitational potential \citep{Grudic2020}. Importantly, the weak lensing data at $g_{\rm bar}\lesssim 10^{-12}-10^{-13}$ m s$^{-2}$ reach beyond the virial radius of the DM halo (Fig.\,\ref{fig:Vc}), so one expects that the data should bend below the relation $g_{\rm obs}=\sqrt{g_\dagger g_{\rm bar}}$ and follow the Newtonian prediction rescaled by the galaxy baryon fraction, $g_{\rm obs}= f_{\rm bar} g_{\rm bar}$. This basic expectation has been confirmed by $\Lambda$CDM simulations \citep{Mercado2024} but is not seen in the lensing data, which instead follows precisely the extrapolation of the kinematic data.

In a MOND context, $g_\dagger$ is identified with $a_0$, which was historically determined using other methods \citep{Milgrom1983b, Milgrom1983a}. MOND predicts that the data should follow the relation $g_{\rm obs}=\sqrt{a_0 g_{\rm bar}}$ down to low accelerations (as observed), as long as galaxies are sufficiently isolated for the external field effect to be negligible \citep{Chae2020, Chae2021}.

\section{Conclusions}

Weak gravitational lensing data allows extending the empirical laws of galaxy dynamics out to very large radii and low accelerations. These empirical laws are difficult to understand in the $\Lambda$CDM context. Remarkably, they were predicted a-priori by MOND.

\vspace{-0.2cm}
\bibliography{GasDynamics}

\begin{thebibliography}{33}
\expandafter\ifx\csname natexlab\endcsname\relax\def\natexlab#1{#1}\fi

\bibitem[{{Begeman}(1989)}]{Begeman1989}
{Begeman}, K.~G. 1989, A\&A, 223, 47

\bibitem[{{Bosma}(1978)}]{Bosma1978}
{Bosma}, A. 1978, PhD thesis, University of Groningen

\bibitem[{{Brouwer} {et~al.}(2021){Brouwer}, {Oman}, {Valentijn}, {Bilicki},
  {Heymans}, {Hoekstra}, {Napolitano}, {Roy}, {Tortora}, {Wright}, {Asgari},
  {van den Busch}, {Dvornik}, {Erben}, {Giblin}, {Graham}, {Hildebrandt},
  {Hopkins}, {Kannawadi}, {Kuijken}, {Liske}, {Shan}, {Tr{\"o}ster},
  {Verlinde}, \& {Visser}}]{Brouwer2021}
{Brouwer}, M.~M., {Oman}, K.~A., {Valentijn}, E.~A., {et~al.} 2021, A\&A, 650,
  A113

\bibitem[{{Buote} \& {Humphrey}(2012)}]{Buote2012}
{Buote}, D.~A. \& {Humphrey}, P.~J. 2012, in Astrophysics and Space Science
  Library, Vol. 378, 235

\bibitem[{{Cappellari}(2016)}]{Cappellari2016}
{Cappellari}, M. 2016, ARA\&A, 54, 597

\bibitem[{{Casertano}(1983)}]{Casertano1983}
{Casertano}, S. 1983, MNRAS, 203, 735

\bibitem[{{Chae} {et~al.}(2021){Chae}, {Desmond}, {Lelli}, {McGaugh}, \&
  {Schombert}}]{Chae2021}
{Chae}, K.-H., {Desmond}, H., {Lelli}, F., {McGaugh}, S.~S., \& {Schombert},
  J.~M. 2021, ApJ, 921, 104

\bibitem[{{Chae} {et~al.}(2020){Chae}, {Lelli}, {Desmond}, {McGaugh}, {Li}, \&
  {Schombert}}]{Chae2020}
{Chae}, K.-H., {Lelli}, F., {Desmond}, H., {et~al.} 2020, ApJ, 904, 51

\bibitem[{{den Heijer} {et~al.}(2015){den Heijer}, {Oosterloo}, {Serra},
  {J{\'o}zsa}, {Kerp}, {Morganti}, {Cappellari}, {Davis}, {Duc}, {Emsellem},
  {Krajnovi{\'c}}, {McDermid}, {Naab}, {Weijmans}, \& {de
  Zeeuw}}]{denHeijer2015}
{den Heijer}, M., {Oosterloo}, T.~A., {Serra}, P., {et~al.} 2015, A\&A, 581,
  A98

\bibitem[{{Desmond}(2017)}]{Desmond2017b}
{Desmond}, H. 2017, MNRAS, 472, L35

\bibitem[{{Desmond} {et~al.}(2019){Desmond}, {Katz}, {Lelli}, \&
  {McGaugh}}]{Desmond2019}
{Desmond}, H., {Katz}, H., {Lelli}, F., \& {McGaugh}, S. 2019, MNRAS, 484, 239

\bibitem[{{Grudi{\'c}} {et~al.}(2020){Grudi{\'c}}, {Boylan-Kolchin},
  {Faucher-Gigu{\`e}re}, \& {Hopkins}}]{Grudic2020}
{Grudi{\'c}}, M.~Y., {Boylan-Kolchin}, M., {Faucher-Gigu{\`e}re}, C.-A., \&
  {Hopkins}, P.~F. 2020, MNRAS, 496, L127

\bibitem[{{Lelli}(2022)}]{Lelli2022}
{Lelli}, F. 2022, Nature Astronomy, 6, 35

\bibitem[{{Lelli} {et~al.}(2013){Lelli}, {Fraternali}, \&
  {Verheijen}}]{Lelli2013}
{Lelli}, F., {Fraternali}, F., \& {Verheijen}, M. 2013, MNRAS, 433, L30

\bibitem[{{Lelli} {et~al.}(2019){Lelli}, {McGaugh}, {Schombert}, {Desmond}, \&
  {Katz}}]{Lelli2019}
{Lelli}, F., {McGaugh}, S.~S., {Schombert}, J.~M., {Desmond}, H., \& {Katz}, H.
  2019, MNRAS, 484, 3267

\bibitem[{{Lelli} {et~al.}(2016){Lelli}, {McGaugh}, {Schombert}, \&
  {Pawlowski}}]{Lelli2016c}
{Lelli}, F., {McGaugh}, S.~S., {Schombert}, J.~M., \& {Pawlowski}, M.~S. 2016,
  ApJ, 827, L19

\bibitem[{{Lelli} {et~al.}(2017){Lelli}, {McGaugh}, {Schombert}, \&
  {Pawlowski}}]{Lelli2017}
{Lelli}, F., {McGaugh}, S.~S., {Schombert}, J.~M., \& {Pawlowski}, M.~S. 2017,
  ApJ, 836, 152

\bibitem[{{McGaugh}(2020)}]{McGaugh2020b}
{McGaugh}, S. 2020, Galaxies, 8, 35

\bibitem[{{McGaugh} {et~al.}(2020){McGaugh}, {Lelli}, {Li}, \&
  {Schombert}}]{McGaugh2020a}
{McGaugh}, S., {Lelli}, F., {Li}, P., \& {Schombert}, J. 2020, in IAU
  Symposium, Vol. 353, Galactic Dynamics in the Era of Large Surveys, ed.
  M.~{Valluri} \& J.~A. {Sellwood}, 144--151

\bibitem[{{McGaugh} {et~al.}(2016){McGaugh}, {Lelli}, \&
  {Schombert}}]{McGaugh2016}
{McGaugh}, S.~S., {Lelli}, F., \& {Schombert}, J.~M. 2016, PRL, 117, 201101

\bibitem[{{McGaugh} {et~al.}(2000){McGaugh}, {Schombert}, {Bothun}, \& {de
  Blok}}]{McGaugh2000}
{McGaugh}, S.~S., {Schombert}, J.~M., {Bothun}, G.~D., \& {de Blok}, W.~J.~G.
  2000, ApJ, 533, L99

\bibitem[{{Mercado} {et~al.}(2024){Mercado}, {Bullock}, {Moreno},
  {Boylan-Kolchin}, {Hopkins}, {Wetzel}, {Faucher-Gigu{\`e}re}, \&
  {Samuel}}]{Mercado2024}
{Mercado}, F.~J., {Bullock}, J.~S., {Moreno}, J., {et~al.} 2024, MNRAS, 530,
  1349

\bibitem[{{Milgrom}(1983{\natexlab{a}})}]{Milgrom1983b}
{Milgrom}, M. 1983{\natexlab{a}}, ApJ, 270, 371

\bibitem[{{Milgrom}(1983{\natexlab{b}})}]{Milgrom1983a}
{Milgrom}, M. 1983{\natexlab{b}}, ApJ, 270, 365

\bibitem[{{Milgrom}(2009)}]{Milgrom2009}
{Milgrom}, M. 2009, ApJ, 698, 1630

\bibitem[{{Milgrom}(2014)}]{Milgrom2014a}
{Milgrom}, M. 2014, MNRAS, 437, 2531

\bibitem[{{Mistele} {et~al.}(2024{\natexlab{a}}){Mistele}, {McGaugh}, {Lelli},
  {Schombert}, \& {Li}}]{Mistele2024b}
{Mistele}, T., {McGaugh}, S., {Lelli}, F., {Schombert}, J., \& {Li}, P.
  2024{\natexlab{a}}, ApJL, 969, L3

\bibitem[{{Mistele} {et~al.}(2024{\natexlab{b}}){Mistele}, {McGaugh}, {Lelli},
  {Schombert}, \& {Li}}]{Mistele2024a}
{Mistele}, T., {McGaugh}, S., {Lelli}, F., {Schombert}, J., \& {Li}, P.
  2024{\natexlab{b}}, JCAP, 2024, 020

\bibitem[{{Noordermeer} {et~al.}(2007){Noordermeer}, {van der Hulst},
  {Sancisi}, {Swaters}, {van Albada}, {dummy}, \& {dummy}}]{Noordermeer2007}
{Noordermeer}, E., {van der Hulst}, J.~M., {Sancisi}, R., {et~al.} 2007, MNRAS,
  376, 1513

\bibitem[{{Rubin} {et~al.}(1978){Rubin}, {Thonnard}, \& {Ford}}]{Rubin1978}
{Rubin}, V.~C., {Thonnard}, N., \& {Ford}, Jr., W.~K. 1978, ApJ, 225, L107

\bibitem[{{Sancisi}(2004)}]{Sancisi2004}
{Sancisi}, R. 2004, in {IAU Symposia}, Vol. 220, Dark Matter in Galaxies, 233

\bibitem[{{Shelest} \& {Lelli}(2020)}]{Shelest2020}
{Shelest}, A. \& {Lelli}, F. 2020, A\&A, 641, A31

\bibitem[{{van Albada} \& {Sancisi}(1986)}]{vanAlbada1986}
{van Albada}, T.~S. \& {Sancisi}, R. 1986, Royal Soc. of London Phil. Transac.
  Series A, 320, 447

\end{thebibliography}
\bibliographystyle{aa}

\end{document}